# The Virtual Laboratory: Enabling On-Demand Drug Design with the World Wide Grid

Rajkumar Buyya[†], Kim Branson[*], Jon Giddy[†,] and David Abramson[†]

[†] School of Computer Science and Software Engg.
Monash University, Caulfield Campus
Melbourne, Australia
*{rajkumar, davida, jon}@csse.monash.edu.au*

[*] Structural Biology
Walter and Eliza Hall Institute
Royal Parade, Parkville, Melbourne
*kbranson@wehi.edu.au*

**Abstract:** Computational Grids are emerging as a popular paradigm for solving large-scale compute and data intensive problems in science, engineering, and commerce. However, application composition, resource management and scheduling in these environments is a complex undertaking. In this paper, we illustrate the creation of a virtual laboratory environment by leveraging existing Grid technologies to enable molecular modeling for drug design on distributed resources. It involves screening millions of molecules of chemical compounds against a protein target, chemical database (CDB) to identify those with potential use for drug design. We have grid-enabled the molecular docking process by composing it as a parameter sweep application using the Nimrod-G tools. We then developed new tools for remote access to molecules in CDB small molecule database. The Nimrod-G resource broker along with molecule CDB data broker is used for scheduling and on-demand processing of jobs on distributed grid resources. The results demonstrate the ease of use and suitability of the Nimrod-G and virtual laboratory tools.

## 1 Introduction

Computational Grids [1] enable sharing a wide variety of geographically distributed resources including supercomputers, storage systems, data sources, and specialized devices owned by different organizations to create virtual enterprises and organizations. They allow selection and aggregation of distributed resources across multiple organizations for solving large-scale computational and data intensive problems in science, engineering, and commerce. The parallel processing of applications on distributed systems provide scalable computing power. This enables exploration of large problems with huge data sets, which is essential for creating new insights into the problem. Molecular modeling for drug design is a scientific application that requires large amounts of computational and data storage capability.

Drug discovery is an extended process that can take as many as 15 years from the first compound synthesis in the laboratory until the therapeutic agent, or drug, is brought to market [11]. Reducing the research timeline in the discovery stage is a key priority for pharmaceutical companies worldwide. Many such companies are trying to achieve this goal through the application and collaboration of advanced technologies such as computational biology, chemistry, computer graphics, and high performance computing (HPC). Molecular modeling has emerged as a popular methodology for drug design—it can combine computational chemistry and computer graphics. Molecular modeling can be implemented as a master-worker parallel application, which can take advantage of HPC technologies such as clusters [2] and grids for large-scale data exploration.

In the context of this paper, molecular modeling involves screening millions of molecules of chemical compounds in a small molecule database, CBD (chemical database) to identify those that are potential drugs. This process is called molecular *docking*. It helps scientists explore how two molecules, such as a drug and an enzyme or protein receptor, fit together (see Figure 1). Docking each molecule in the target chemical database is both a compute and data intensive task. It is our goal to use Grid technologies to provide cheap and efficient solutions to the modeling process.

While performing docking, information about the molecule must be extracted from one of a number of large chemical databases. Because the databases require storage space in the order of hundreds of megabytes to terabytes, it isn't feasible to transfer the chemical database to all nodes in the Grid while processing. Therefore, access to chemical database must be provided as *network service* (see Figure 2).

Also, the database needs to be selectively replicated on a few nodes within the Grid to avoid any bottleneck due to having chemical database serving from a single source. Intelligent mechanisms (e.g., CBD broker) need to be supported for selecting optimal source for CDB services depending a resource on which docking job is to be processed.

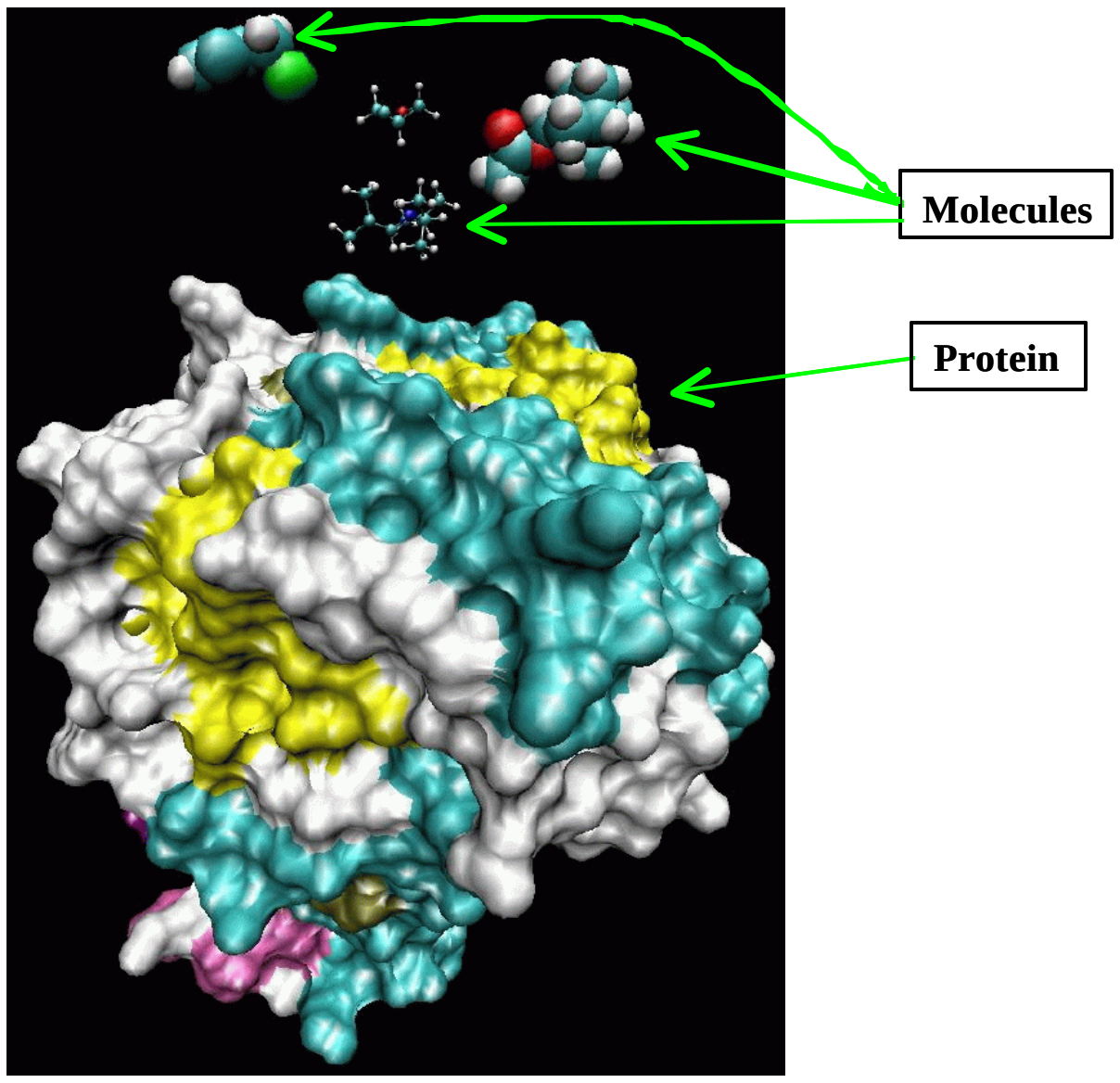

**Figure 1:** The x-ray crystal structure of a target receptor (surface rendered protein) and small molecules binding.

As part of the Virtual Laboratory project, we have investigated the requirements of running the molecular modeling application on the Grid. We have leveraged existing Grid technologies and developed new tools that are essential for Grid enabling the chemical database, and the docking application on distributed resources. In this paper, we discuss a layered architecture of technologies and tools for creating the virtual laboratory environment for drug design application. We present results of scheduling molecular docking application for processing on the WWG (world wide grid) testbed along with conclusion.

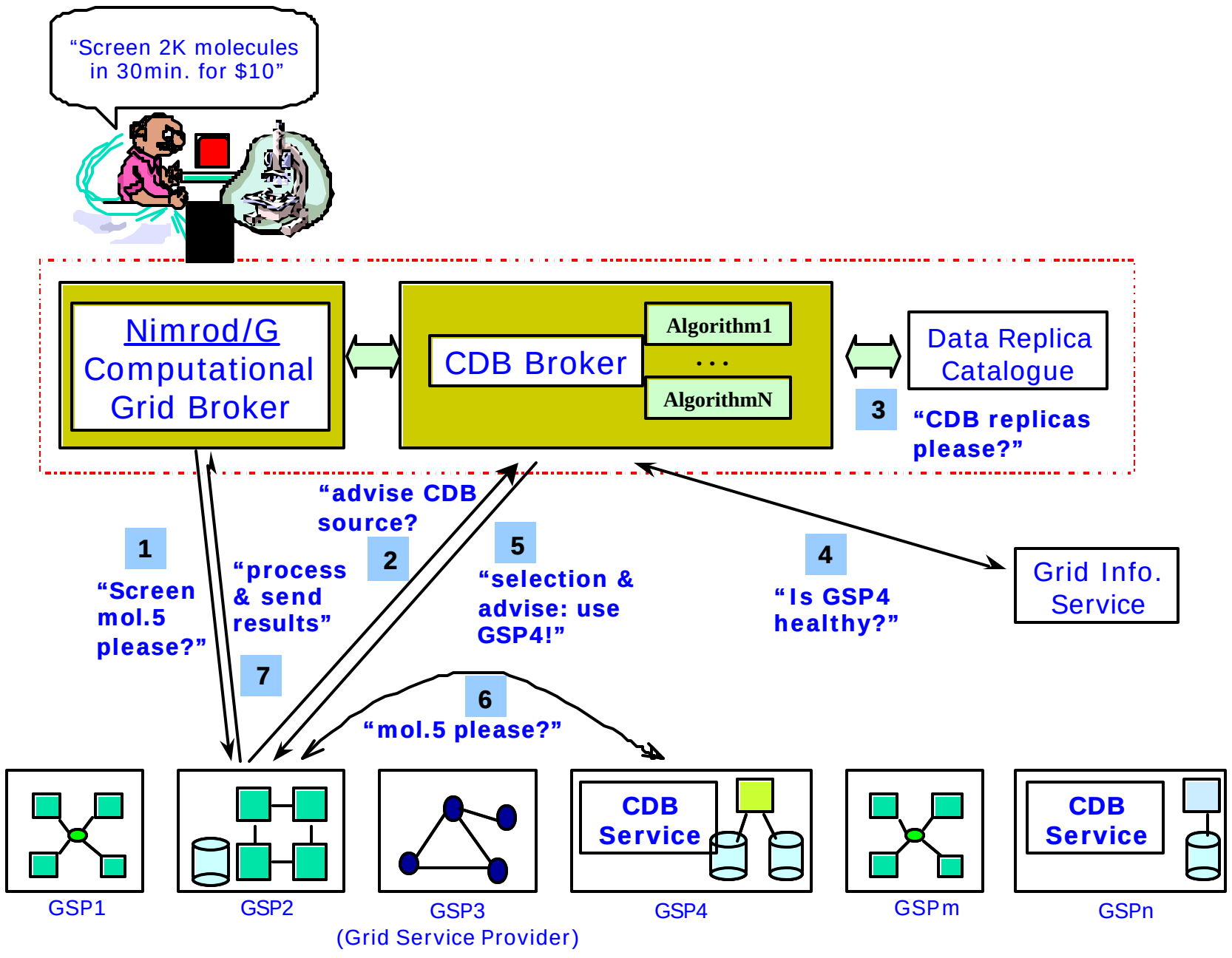

**Figure 2:** Resource brokering architecture for screening molecules on distributed resources.

## 2 Architecture – The Software Stack

The virtual laboratory builds on the existing Grid technologies and tools for performing data intensive computing on distributed resources. To provide a virtual computing environment for applications, depending on their requirements, we have developed necessary tools. For example, to enable drug design application's docking for on large databases, we have developed tools for providing access to CDB databases as network service. Many applications (e.g., molecular modeling, high-energy physics events processing, and financial investment risk-analysis) explore range of scenarios. Instead of explicitly developing them as parallel application using interfaces such as MPI, they can be composed as parameter sweep applications using tools such as Nimrod. Such application jobs can be executed in parallel on distributed resources using the Nimrod-G resource broker (see Figure 2). A layered architecture and the software stack essential for performing molecular modeling on distributed resources is depicted in Figure 3. The Grid technologies, tools, and application software used are discussed below:

- The DOCK software for Molecular Modeling [14].
- The Nimrod Parameter Modeling Tools [3][16] for enabling DOCK as parameter sweep application.
- The Nimrod-G Grid Resource Broker [4][5] for scheduling DOCK jobs on the Grid.
- Chemical Database (CDB) Management and Intelligent Access Tools:
    - CDB database Lookup/Index Table Generation.
    - CDB and associated index-table Replication.
    - CDB Replica Catalogue for CDB resource discovery.
    - CDB Servers for providing CDB services
    - CDB Brokering for selecting suitable (Replica Selection).
    - CDB Clients for fetching Molecule Record (Data Movement).
- The GrACE for resource trading toolkit [6].
- The Globus middleware for secure and uniform access to distributed resources [9].

The Grid fabric multiprocessor or cluster resources are managed as single entity using resource management systems such as OS-fork, LSF, Condor, and SGE.

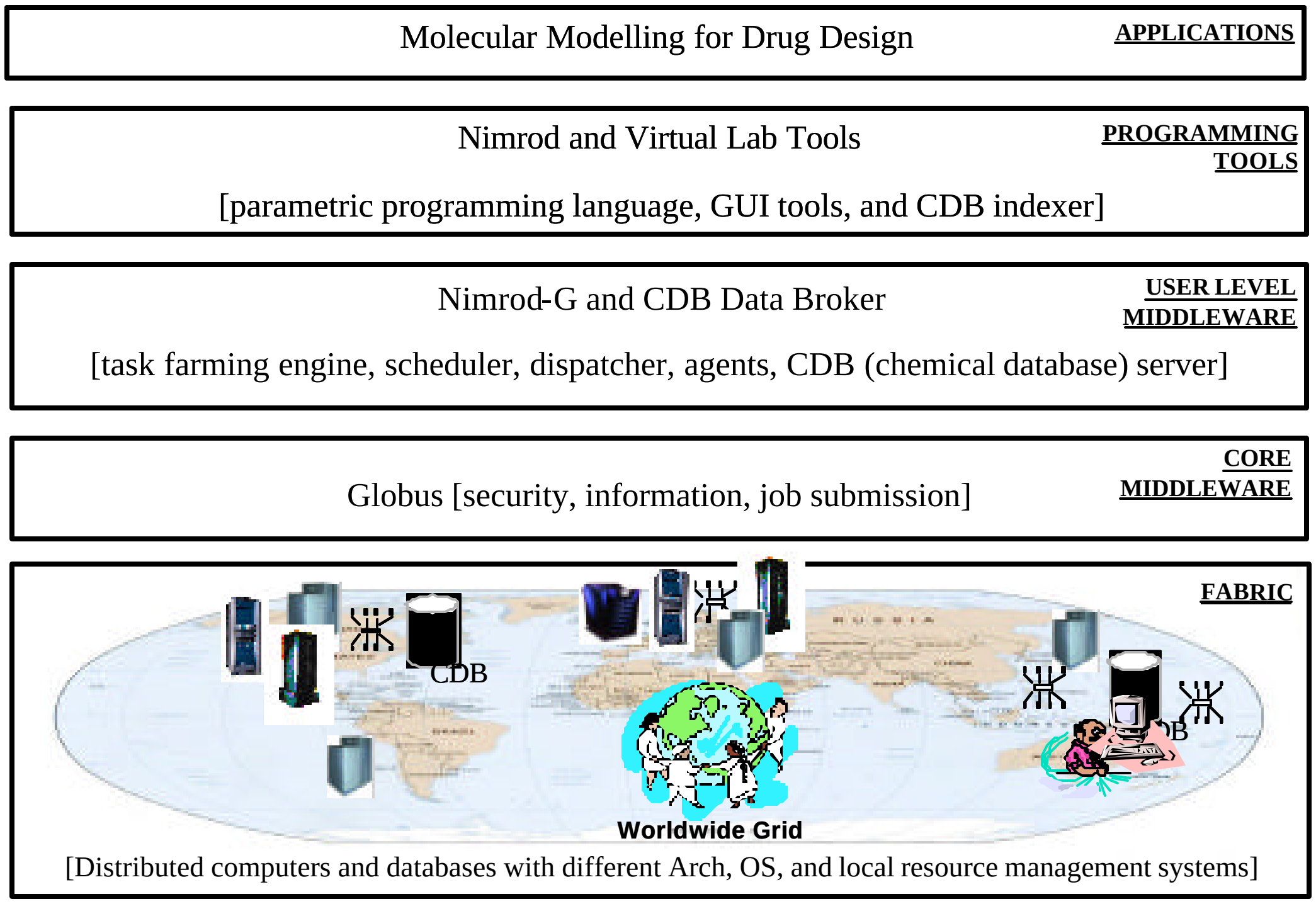

**Figure 3:** Layered architecture of Virtual Lab for Drug Design.

## 2.1 Docking Code

The original docking code developed by researchers at the University of California in San Francisco (UCSF) is one of the most popular molecular docking applications [10]. The program evaluates the chemical and geometric complementarities between a small molecule and a macromolecular binding site. It explores ways in which two molecules, such as a drug and an enzyme or protein receptor, might fit together. Compounds that might bind tightly to the target receptor must have complementary chemical and spatial natures. Thus docking can be seen as a 3 dimensional puzzle searching for pieces that will fit into the receptor site. It is important to be able to identify small molecules (compounds), which may bind to a target macromolecule. This is because *a compound, which binds to a biological macromolecule, may modulate its function, and thus may with further development eventually become a drug.* An example of such a drug is the anti influenza drug Relenza which functions by binding to influenza virus attachment proteins thus preventing viral infection.

The relationship between the key programs in the dock suite is depicted in Figure 4. The receptor coordinates at the top represent 3D structure of protein. The molecular modeller identifies the active site, and other sites of interest, and uses the program "sphgen" to generate the sphere centers, which fill the site [13]. The program "grid" generates the scoring grids [14]. The program "dock" matches spheres (generated by sphgen) with ligand atoms and uses scoring grids (from grid) to evaluate ligand orientations [13] [14]. It also minimizes energy-based scores [15] [6]. The focus of our work is on docking molecules in CDB with receptor to identify potential compounds that can act as "drug". Hence, discussion in this paper is centered on the execution of the program "dock" as parameter sweep application on world-wide distributed resources.

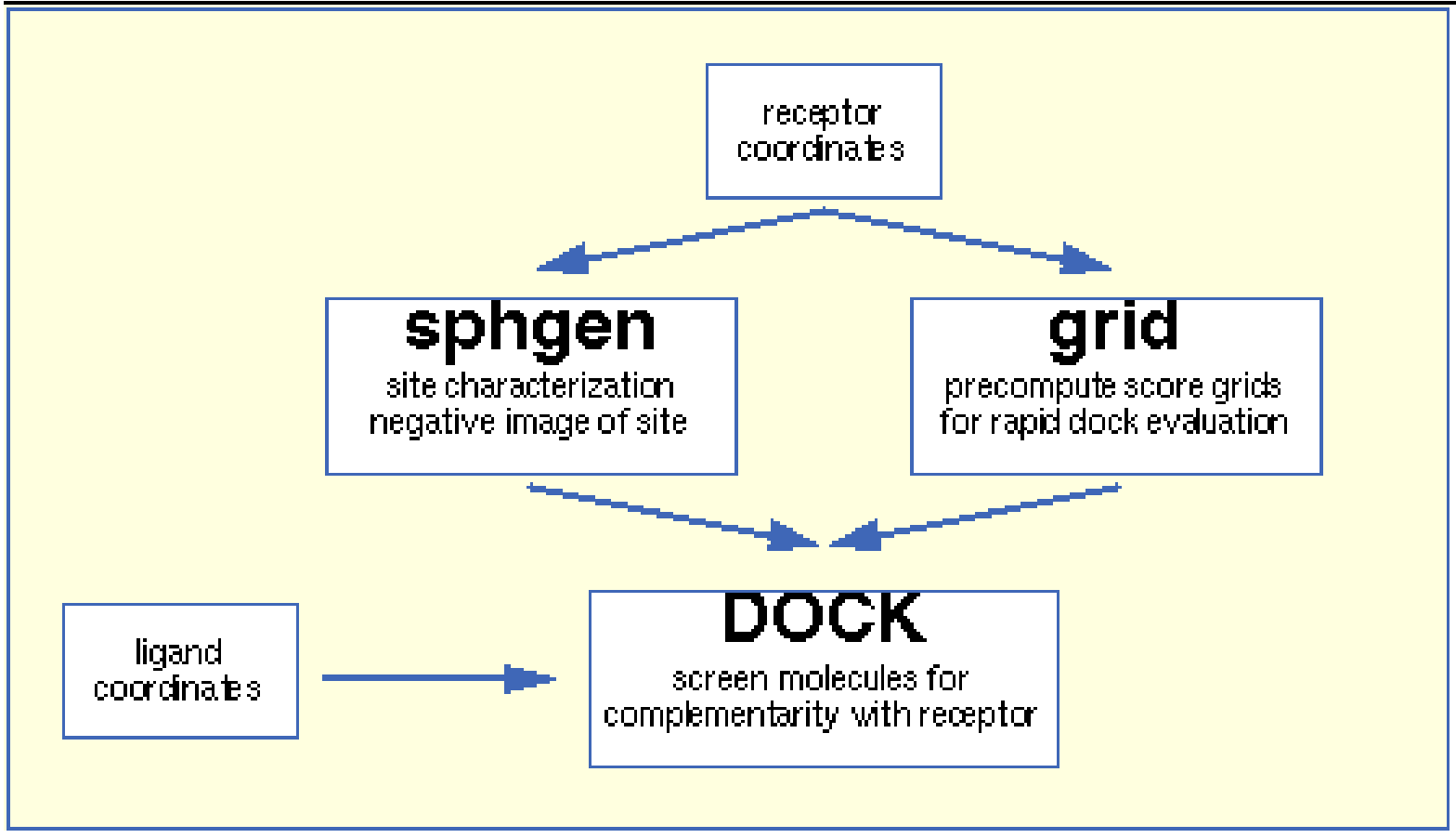

**Figure 4:** Relation between key programs in the dock suite [10].

The docking code is highly portable. We have been able to produce executables for Sun-Solaris, PC Linux, SGI IRIX, and Compaq Alpha/OSF1 architectures. For docking on heterogeneous resources, the Nimrod-G broker selects the right executable depending on resource architecture.

## 2.2 Nimrod-G Tools

Nimrod-G provides a suite of tools and services for creating parameter sweep applications, performing resource management, and scheduling applications. They are, a simple declarative programming language and associated GUI tools for creating scripts and parameterization of application input data files, and a grid resource broker with programmable entities for processing jobs on grid resources.

***Tools for Creating Parameter Sweep Applications***

Nimrod supports declarative programming language and GUI-based tools that assist in the creation of parameter sweep applications [3]. They allow user to: a) parameterise input files, b) prepare a plan file containing the commands that define parameters and their values c) generate a run file, which converts the generic plan file to a detailed list of jobs; d) schedule jobs for processing on distributed machines, and e) control and monitor execution of the jobs. The application execution environment handles online creation of input files and command line arguments through parameter substitution. The GUI tools supported by enFuzion, a commercial version of Nimrod, can also be used for parameterising applications. enFusion uses the same syntax as Nimrod, details on syntax of parametric language in the Enfuzion user manual [16]. Both Nimrod and enFuzion have been successfully used for performing parameter studies in a single administrative domain such as clusters. Nimrod-G [4][5] extends the capabilities of Nimrod and EnFuzion with the addition of powerful resource discovery, trading, scheduling algorithms [7]. In Section 3, we discuss capabilities of Nimrod tools by composing a molecular modeling program as a parameter sweep application for docking compounds in CDB databases and processing docking jobs on the Grid.

***Nimrod-G Grid Resource Broker for scheduling DOCK jobs on Grid***

The Nimrod-G Resource broker is responsible for determining the specific requirements that an experiment places on the Grid and performing resource discovery, scheduling, dispatching jobs to remote Grid nodes, starting and managing job execution, and gathering results back to the home node [5]. The sub-modules of our resource broker are, the task farming engine; the scheduler, a schedule advisor backed with scheduling algorithms, and a resource trading manager; a dispatcher and actuators for deploying agents on grid resources; and agents for managing execution of Nimrod-G jobs on grid resources.

The Nimrod-G grid explorer and dispatcher components are implemented using Globus services for resource discovery and starting execution on Nimrod-G agents, which then takes care of all operations related to execution of an assigned job. The interaction between components of the Nimrod-G runtime machinery and Grid services during runtime is shown in Figure 5. The Nimrod-G broker supports *deadline and budget constrained (DBC) scheduling* algorithms driven by a computational economy and user requirements [7]. In section 4, we discuss the results of Nimrod-G broker scheduling molecular modeling application on the Grid with DBC time and cost optimization scheduling algorithms.

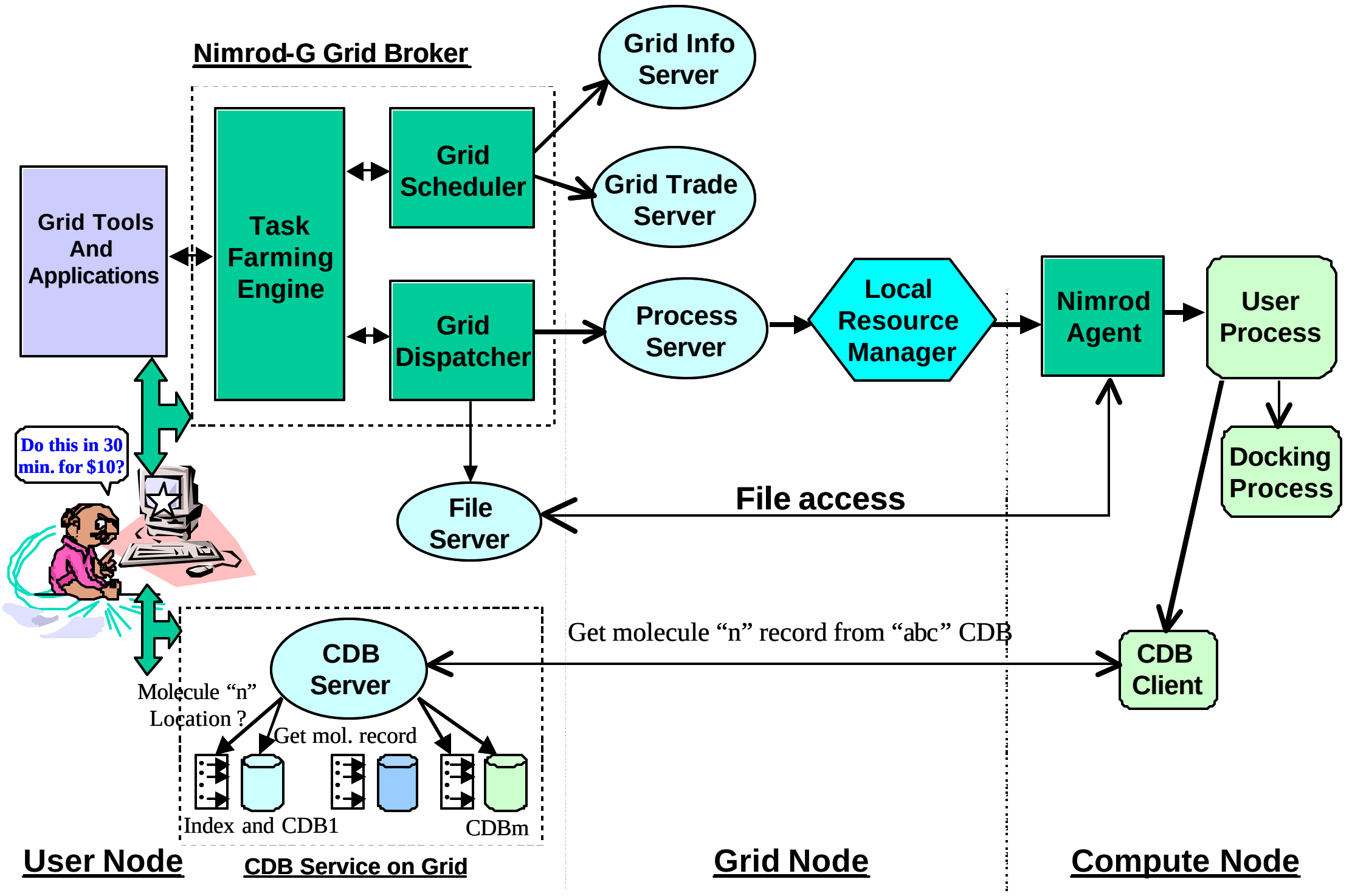

**Figure 5:** Layered architecture of Virtual Lab for Drug Design.

## 2.3 Chemical Database (CDB) Management and Intelligent Access Tools

The CDB databases consist of a large number of small molecules from commercially available organic synthesis libraries, and natural product databases. The molecules are represented in MOL2 file (.mol2) format [12], which is a portable representation of a SYBYL molecule. The MOL2 file is an ASCII file which contains all the information needed to reconstruct a SYBYL molecule. Each molecule record in CDB represents the three-dimensional (3D) structural information of a compound. The numbers of compounds in each CDB can be in the order of tens of thousands and the database size be anywhere from tens of Megabytes to Gigabytes and even Terabytes.

Instead of replicating the CDB database on every Grid node, we have developed a mechanism for replicating it on only a few selected nodes and then accessing them on demand from remote machines. The CDB molecule server needs to provide services to multiple users for molecule information from any request database. Hence, we have developed a multithreaded CDB server that can service requests from multiple users simultaneously. Instead of searching molecule records sequentially in a database represented using MOL2 format, we have tools for creating index-tables for each CDB along with record size information. When a molecule record is requested, a CDB server first looks at CDB index file to identify record location and size and then directly reads the record from the CDB file. An interaction between a Grid node and a node running CDB server while performing docking is shown in Figure 5.

It is possible to screen virtual combinatorial databases, in their entirety. This methodology allows only the required compounds to be subjected to physical screening and/or synthesis reducing both time and expense.

## 3 Application Composition

Nimrod/G experiments are configured from a plan file written using a simple declarative language. The plan file specifies the parametric tasks and the types of the parameters to these tasks. A parametric task consists of a script defined using a sequence of simple commands, providing the ability to copy files to and from the remote node, execute certain programs, and perform parameter substitutions in input files.

Figure 6 shows an input file for the molecular docking program. This configures several aspects of the application, including the "ligand_atom_file" attribute, which indicates the molecule number in CDB to be docked. Nimrod/G will sweep over a range of molecules, so for each molecule it needs to place a different name at this position in the input file. To accomplish this, the user must replace the current value, naming a particular file, with a substitution placemarker. A substitution placemarker appears familiar to most users of Unix shells. It consists of a dollar-sign ($) followed by the name of the parameter controlling the substitution, optionally surrounded by braces.

Figure 7 shows the input file with several attributes replaced by substitution place markers. The first part of the "ligand_atom_file" attribute has been replaced by a place marker for the parameter "ligand_number".

Figure 8 shows the parameter definition section of the plan file. Each parameter is defined by a keyword "parameter", followed by the parameter name, an optional label, and a parameter type. The remaining information on each line defines valid values for the parameter.

The parameter, for example, "database_name" has a label, and is of type text. Its valid values are listed, and the user will be able to select one of the values for the duration of the entire experiment. Most of the remaining parameters are single values, either text strings or integers, selected by the user, but with default values provided if the user does not wish to choose a value.

The parameter, "ligand_number", used to select the molecule, is defined as an integer, which can range from 1 to 2000.

The plan file is typically provided as input to a job generation tool, such as the TurboLinux EnFuzion Generator, in order to create a run file. The run file is similar to a plan file but contains specific instances of jobs to be run.

The parameters “receptor_site_file” and “score_grid_prefix” indicate the data input files. Their values indicate that data input files are located in the user home directory on Grid nodes. Instead of pre-staging, these files can be copied at runtime by defining necessary “copy” operation in the job’s “nodestart” or “main” task (see Figure 9). It is advisable to “pre-stage” large input files if they are going to be used for docking with many databases.

During the conversion, the user is able to set concrete values for each of the parameters. For the parameter "ligand_number", the user may choose not to select all values from 1 to 2000, but may select a subset of these values. By default, this generated 2000 jobs, each docking a single molecule.

The run file then contains a job for each combination of parameters. Hence the number of jobs is the product of the number of values chosen for each parameter. Since most of the parameters except "ligand_number" are single-valued, they have no effect on the number of jobs.

Figure 9 shows the task definition section of the plan file. The "nodestart" task is performed once for each remote node. Following that, the files copied during that stage are available to each job when it is started. The "main" task controls the actions performed for each job.

```
score_ligand                   yes
minimize_ligand                yes
multiple_ligands               no
random_seed                    7
anchor_search                  no
torsion_drive                  yes
clash_overlap                  0.5
conformation_cutoff_factor     3
torsion_minimize               yes
match_receptor_sites           no
random_search                  yes
    . . . . . .
    . . . . . .
maximum_cycles                 1
ligand_atom_file               S_1.mol2
receptor_site_file             ece.sph
score_grid_prefix              ece
vdw_definition_file            parameter/vdw.defn
chemical_definition_file       parameter/chem.defn
chemical_score_file            parameter/chem_score.tbl
flex_definition_file           parameter/flex.defn
flex_drive_file                parameter/flex_drive.tbl
ligand_contact_file            dock_cnt.mol2
ligand_chemical_file           dock_chm.mol2
ligand_energy_file             dock_nrg.mol2
```

Molecule to be screened

**Figure 6:** A configuration input file for docking application.

```
score_ligand                   $score_ligand
minimize_ligand                $minimize_ligand
multiple_ligands               $multiple_ligands
random_seed                    $random_seed
anchor_search                  $anchor_search
torsion_drive                  $torsion_drive
clash_overlap                  $clash_overlap
conformation_cutoff_factor     $conformation_cutoff_factor
torsion_minimize               $torsion_minimize
match_receptor_sites           $match_receptor_sites
random_search                  $random_search
    . . . . . .
    . . . . . .
maximum_cycles                 $maximum_cycles
ligand_atom_file               ${ligand_number}.mol2
receptor_site_file             $HOME/dock_inputs/${receptor_site_file}
score_grid_prefix              $HOME/dock_inputs/${score_grid_prefix}
vdw_definition_file            vdw.defn
chemical_definition_file       chem.defn
chemical_score_file            chem_score.tbl
flex_definition_file           flex.defn
flex_drive_file                flex_drive.tbl
ligand_contact_file            dock_cnt.mol2
ligand_chemical_file           dock_chm.mol2
ligand_energy_file             dock_nrg.mol2
```

Molecule to be screened

**Figure 7:** Parameterisation of a configuration input file.

```
parameter database_name label "database_name" text select oneof "aldrich"
    "maybridge" "maybridge_300" "asinex_egc" "asinex_epc" "asinex_pre"
    "available_chemicals_directory" "inter_bioscreen_s" "inter_bioscreen_n"
    "inter_bioscreen_n_300" "inter_bioscreen_n_500" "biomolecular_research_institute"
    "molecular_science" "molecular_diversity_preservation"
    "national_cancer_institute" "IGF_HITS" "aldrich_300" "molecular_science_500"
    "APP" "ECE" default "aldrich_300";
parameter CDB_SERVER text default "bezek.dstc.monash.edu.au";
parameter CDB_PORT_NO text default "5001";parameter score_ligand text default "yes";
parameter minimize_ligand text default "yes";
parameter multiple_ligands text default "no";
parameter random_seed integer default 7;
parameter anchor_search text default "no";
parameter torsion_drive text default "yes";
parameter clash_overlap float default 0.5;
parameter conformation_cutoff_factor integer default 5;
parameter torsion_minimize text default "yes";
parameter match_receptor_sites text default "no";
parameter random_search text default "yes";
    . . . . . .
    . . . . . .
parameter maximum_cycles integer default 1;
parameter receptor_site_file text default "ece.sph";
parameter score_grid_prefix text default "ece";
parameter ligand_number integer range from 1 to 2000 step 1;
```

**Molecules to be screened**

**Figure 8:** A plan file defining parameters type and their values.

```
task nodestart
      copy ./parameter/vdw.defn node:.
      copy ./parameter/chem.defn node:.
      copy ./parameter/chem_score.tbl node:.
      copy ./parameter/flex.defn node:.
      copy ./parameter/flex_drive.tbl node:.
      copy ./dock_inputs/get_molecule node:.
      copy ./dock_inputs/dock_base node:.
endtask
task main
           node:substitute dock_base dock_run
           node:substitute get_molecule get_molecule_fetch
           node:execute sh ./get_molecule_fetch
           node:execute $HOME/bin/dock.$OS -i dock_run -o dock_out
           copy node:dock_out ./results/dock_out.$jobname
           copy node:dock_cnt.mol2 ./results/dock_cnt.mol2.$jobname
           copy node:dock_chm.mol2 ./results/dock_chm.mol2.$jobname
           copy node:dock_nrg.mol2 ./results/dock_nrg.mol2.$jobname
endtask
```

**Figure 9:** Task definition of docking jobs.

The first line of the "main" task performs parameter substitution on the file "dock_base", creating a file "dock_run". This is the action that replaces the substitution place markers in our input file with the actual values for the job.

As each docking operation is performed on a selected molecule in the CDB database, it is not necessary to copy such large databases on all Grid nodes. Hence, not only is the molecule file named in the configuration file, we also go to particular lengths to copy only the data for the molecule being tested. The executable script "get_molecule_fetch" (see Figure 10) is also created using parameter substitution, and runs the "mmclient" executable, which fetches the molecule record from the *CDB molecule server* based on the parameter "ligand_number". The molecule record is saved in a file whose name is the same as integer value of the “ligand_number” parameter and “mol2” as its extension. For instance, if the parameter ligand_number value is 5, then molecule record will be saved in a file “5.mol2”.

```
#!/bin/sh
$HOME/bin/mmclient.$OS $PDB_SERVER $PDB_PORT_NO ${database_name}.db $ligand_number
```

**Figure 10**: Parameterisation of script for extracting molecule from PDB.

The main code is the "dock" executable. Note that in the “execute” command, there are pseudo-parameters that do not appear in the plan file. These include environment variables, such as "HOME", as well as other useful parameters, such as "OS" indicating the operating system on the node. This allows us to select the correct executable for the node. If the “dock” executable files do not exist on Grid nodes, they need to be copied at runtime as part of the job’s “nodestart” task similar to copying input files.

The dock_run file created in the substitution step previously is now provided as the input configuration file for the docking process. The output files are then copied back to the local host, and renamed with another pseudo-parameter, the unique "jobname" parameter.

# 4 Experimentation

We have performed scheduling experiments from a grid resource in Australia along with four resources available in Japan and one in USA. Table 1 shows the site of resources and their properties, Grid services and access price (G$/CPU second). We have performed a trial screening 200 molecules (from aldrich_300 CDB) on a target enzyme (protein) called endothelin converting enzyme (ECE), which is involved in hypotension. The receptor 3D structure is derived from homology modeling using related receptor structures whose three dimensional structures have been solved by x-ray crystallography experiments. Currently we are processing tens of thousands of molecules on resources in the WWG testbed [17]. Those results will be presented in the final paper to replace the ones presented here.

| **Organization & Location** | **Vendor, Resource Type, # CPU, OS, hostname** | **Grid Services and Fabric, Role** | **Price (G$/CPU sec.)** | **Number of Jobs Executed** | |
|---|---|---|---|---|---|
| | | | | **TimeOpt** | **CostOpt** |
| Monash University, Melbourne, Australia | Sun: Ultra-1, 1 node, bezek.dstc.monash.edu.au | Globus, Nimrod-G, CDB Server, Fork (Master node) | -- | -- | -- |
| AIST, Tokyo, Japan | Sun: Ultra-4, 4 nodes, Solaris, hpc420.hpcc.jp | Globus, GTS, Fork (Worker node) | 1 | 44 | 102 |
| AIST, Tokyo, Japan | Sun: Ultra-4, 4 nodes, Solaris, hpc420-1.hpcc.jp | Globus, GTS, Fork (Worker node) | 2 | 41 | 41 |
| AIST, Tokyo, Japan | Sun: Ultra-4, 4 nodes, Solaris, hpc420-2.hpcc.jp | Globus, GTS, Fork (Worker node) | 1 | 42 | 39 |
| AIST, Tokyo, Japan | Sun: Ultra-2, 2 nodes, Solaris, hpc220-2.hpcc.jp | Globus, GTS, Fork (Worker node) | 3 | 11 | 4 |
| Argonne National Lab, Chicago, USA | Sun: Ultra -8, 8 nodes, Solaris, pitcairn.mcs.anl.gov | Globus, GTS, Fork (Worker node) | 1 | 62 | 14 |
| | | Total Experiment Cost (G$) | | 17702 | 14277 |
| | | Time to Finish Experiment (Min.) | | 34 | 59.30 |

**Table 1**: The WWG testbed resources used in scheduling experiments, job execution and costing.

In this scheduling experiment, we have set 60 minutes as the deadline limit and 50,000 G$ (grid dollar) as the budget limit. We conducted experiments for two different optimization strategies:

1. Optimize for Time - this strategy produces results as early as possible, but before a deadline and

within a budget limit.

2. Optimize for Cost - this strategy produces results by deadline, but reduces cost within a budget limit.

The first experiment, *Optimize for Time* scheduling, was performed on November 3, 2001 at 23:23:00, Australian Eastern Standard Time (A EST), with 60 minutes deadline and finish on November 3, 2001 by 23:57:00. It took 34 minutes to finish the experiment using resources available at that time with the expense of 17,702 G$. Figure 11 shows number of jobs processed on different resources selected depending on their cost and availability. Figure 12 shows corresponding expenses of processing so far on resources. Figure 13 shows the number of jobs in execution on resources at different times. From the graphs, it can be observed that the broker selected resources to ensure that the experiment is completed at the earliest possible time given the current availability of resources and the budget limitations. After 30 minutes, it discovered that it can still complete early without using the most expensive resource, "hpc220-2.hpcc.jp".

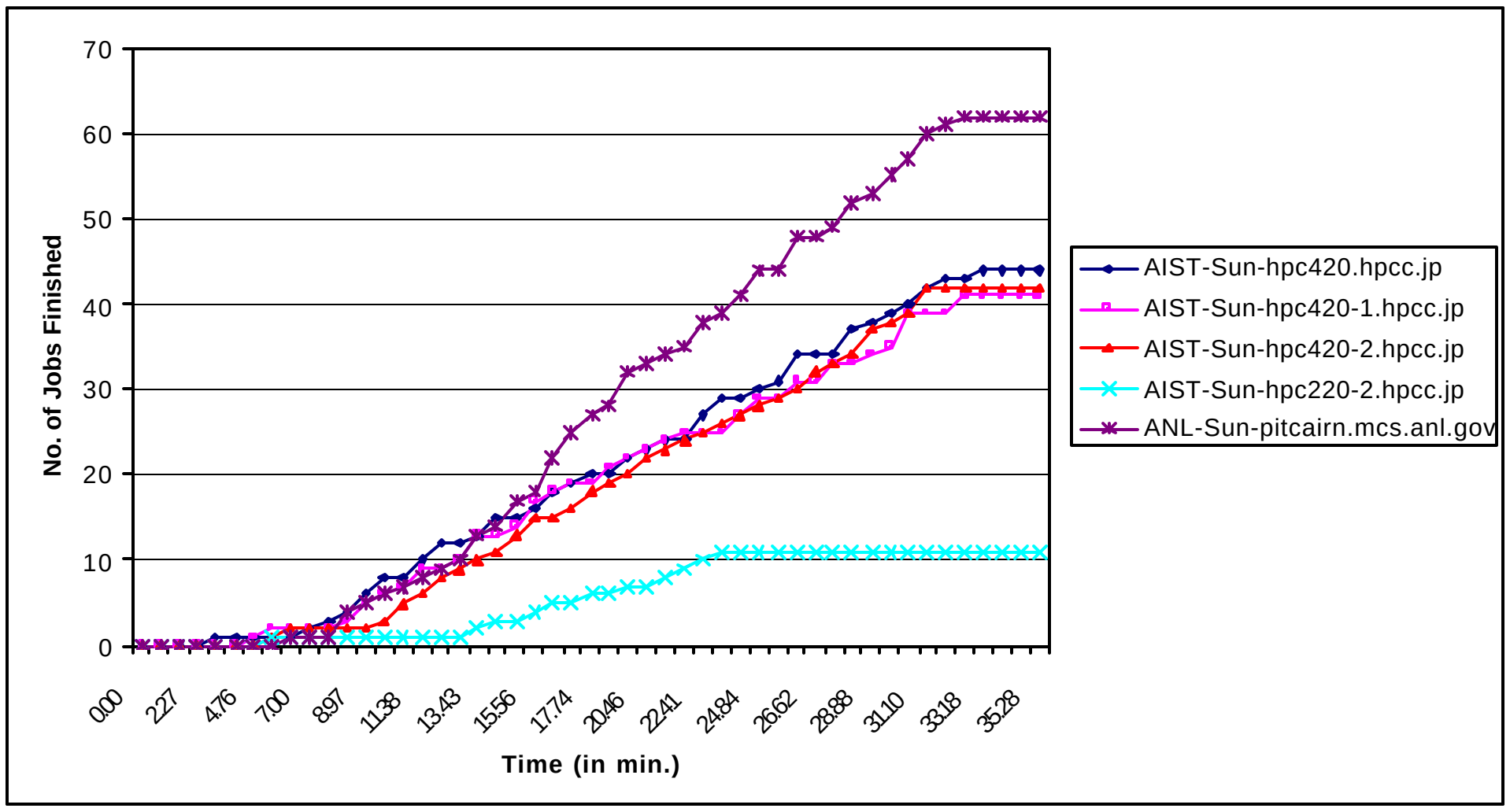

**Figure 11:** No. of jobs processed on Grid resources during DBC time optimization scheduling.

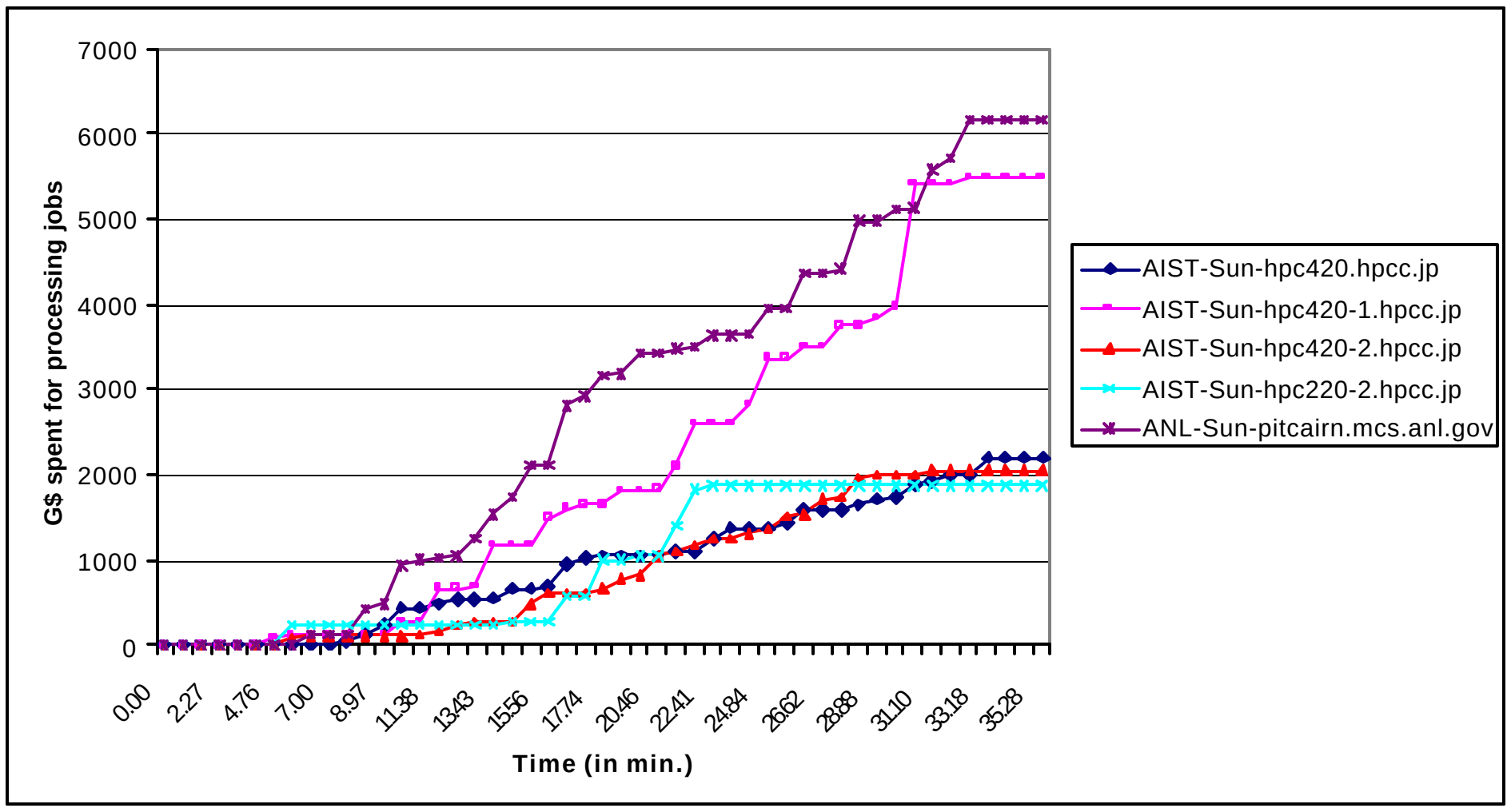

**Figure 12:** The budget amount spent on resources during DBC time optimization scheduling.

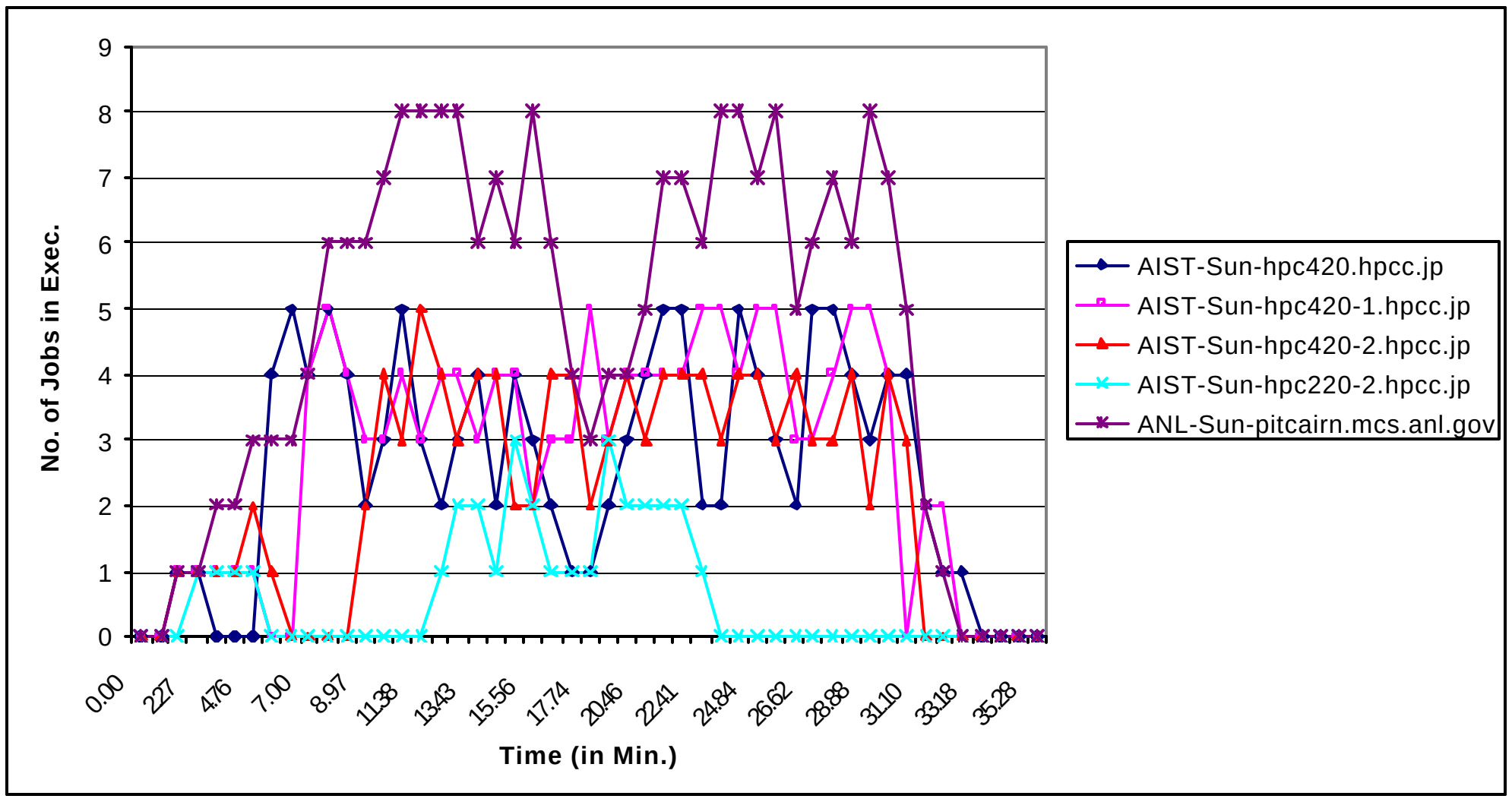

**Figure 13:** No. of jobs in execution on Grid resources during DBC time optimization scheduling.

The second experiment, *Optimize for Cost* scheduling, was performed on November 4, 2001 at 00:08:00, AEST, with 60 minutes deadline and finish on November 4, 2001 by 01:07:30. It took almost 59.30 minutes to finish the experiment using resources available at that time with the expense of 14,277 G$. It is interesting to note that the second experiment took extra 25.40 minutes, but saved 3,475 G$ in the process. Figure 14 shows number of jobs processed on different resources selected depending on their cost and availability. Figure 15 shows corresponding expenses of processing so far on resources. Figure 16 shows the number of jobs in execution on resources at different times. From the graphs, it can be observed that the broker selected cheapest resources to ensure that the experiment is completed with minimum expenses, but on or before the deadline limit. In the beginning expensive resources are used to ensure that the deadline can be meet. If for any reason cheapest resources are unable to deliver expected performance, then the broker seeks the help of expensive resources to meet the deadline.

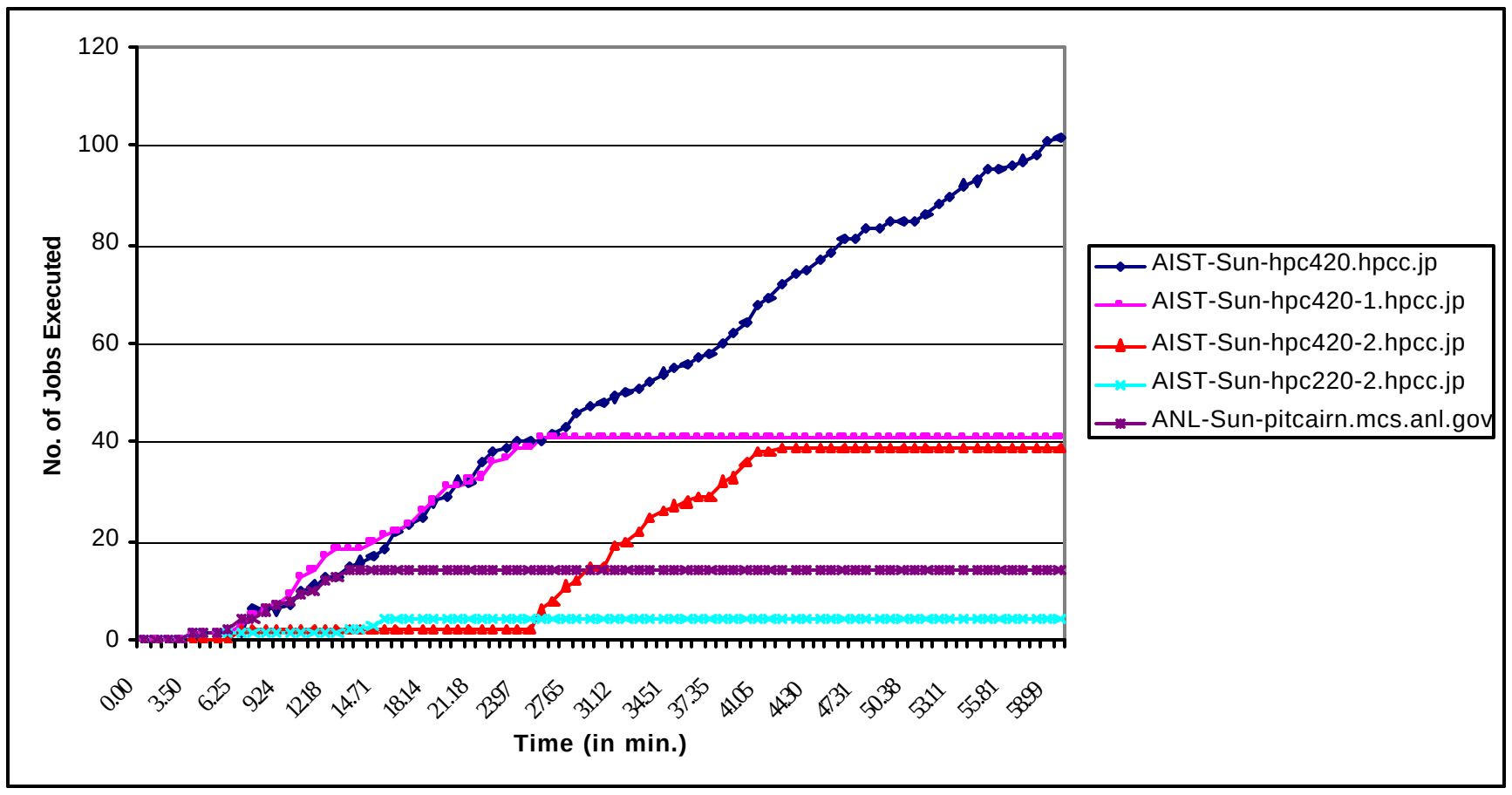

**Figure 14:** No. of jobs processed on Grid resources during DBC Cost optimization scheduling.

## 5 Conclusion and Future Work

Computational Grids enable distributed resource sharing and aggregation for solving large scientific problems. However, application composition, data management, and scheduling in this environment is a complex undertaking. We have developed a virtual laboratory environment tools for processing molecular modeling applications on the Grid by leveraging existing Grid technologies. The new tools developed include chemical database indexer, CBD server for providing access to molecules in CDB databases as a network service, clients for accessing CBD services from a selected CBD service. The Nimrod-G toolkit provides simple and easy to use tools for composing an existing application as a parameter sweep application and processing them on distributed resources. The results of Nimrod-G broker scheduling molecular docking application jobs on distributed Grid resources show its potential for service oriented computing. Computational economy and quality of services (QoS) driven scheduling systems enable users to utilize resources effectively by trading off between deadline and budget depending on their QoS requirements. From our experience with developing a prototype of virtual laboratory environment for distributed drug design shows the potential and applicability of Grid and Nimrod-G tools for data intensive computing.

We are extending the current system to support adaptive mechanism for selection of the best CDB service depending access speed and cost. We are also looking into applying experience gained in this work to develop virtual laboratory environment for enabling high-energy physics' events processing on distributed resources in larger scale.

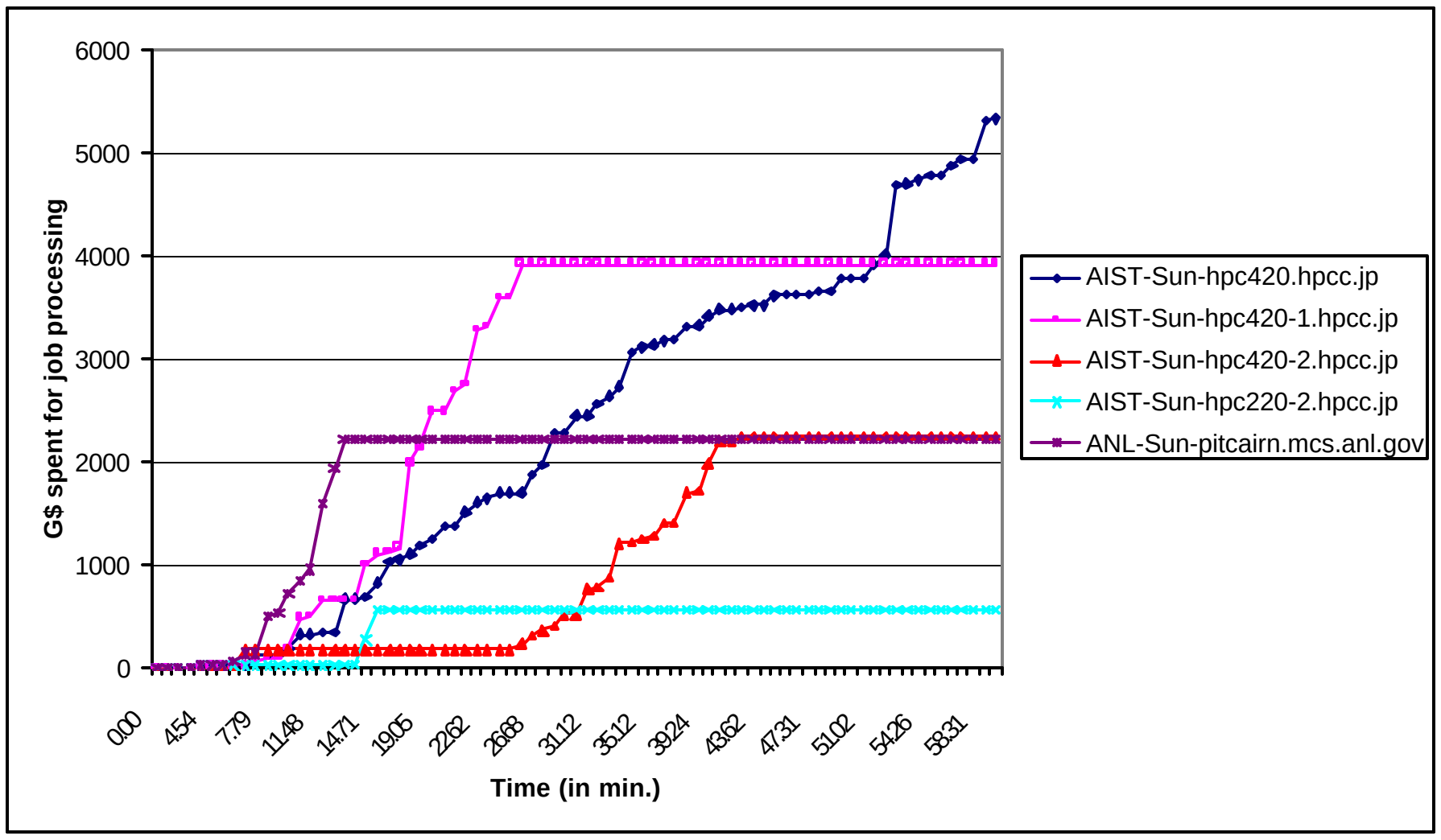

**Figure 15:** The budget amount spent on resources during DBC Cost optimization scheduling.

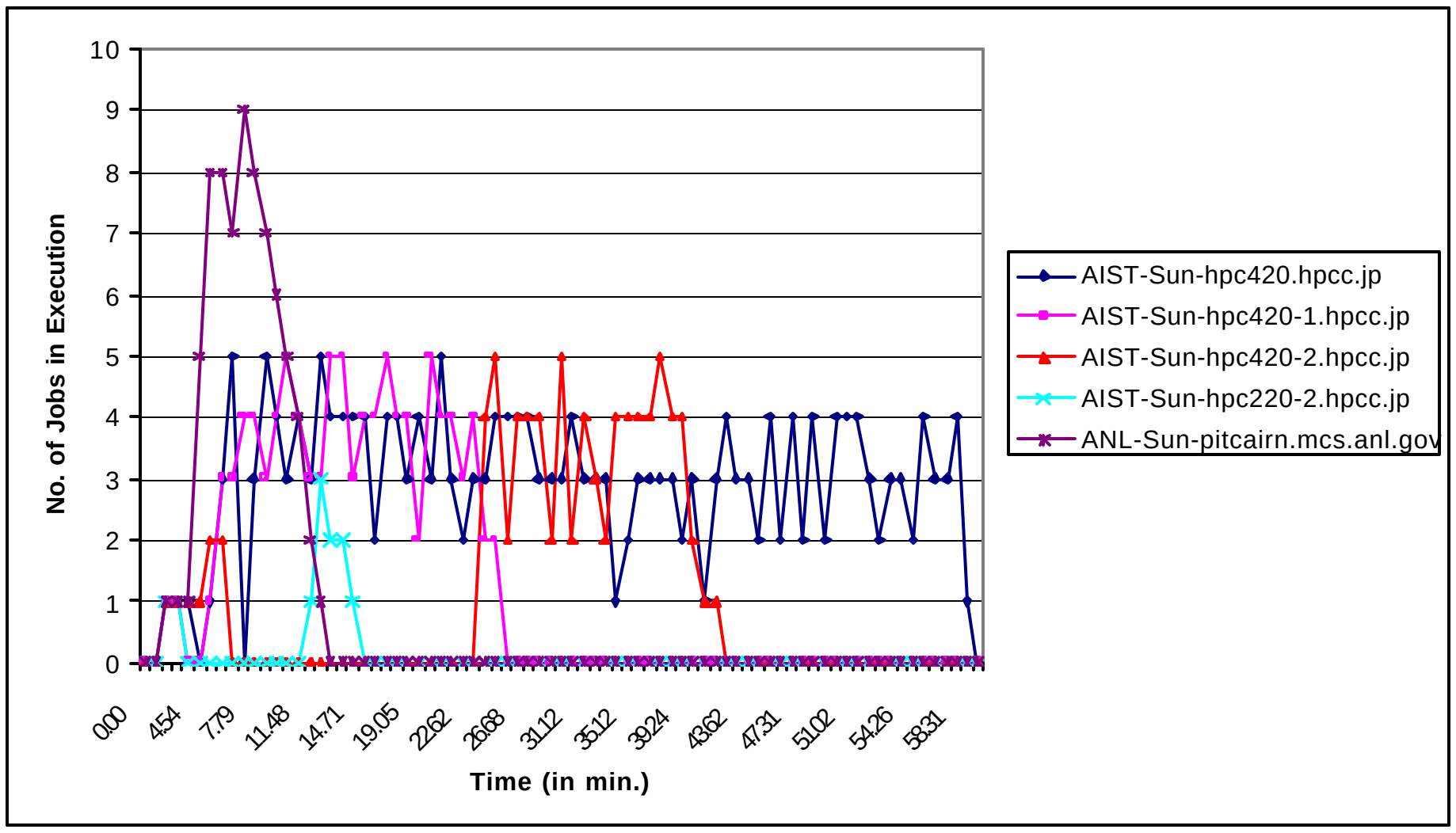

**Figure 16:** No. of jobs in execution on Grid resources during DBC Cost optimization scheduling.

## Acknowledgements

We would like to thank Rob Gray for his comments on improving the paper. We thank National Institute of Advanced Industrial Science and Technology, Tokyo, Japan and the Argonne National Laboratory,

Chicago, USA for providing access to the resources used in the experimentation.

## References

[1] I. Foster and C. Kesselman (editors), *The Grid: Blueprint for a Future Computing Infrastructure*, Morgan Kaufmann Publishers, USA, 1999.

[2] R. Buyya (editor), High Performance Cluster Computing: Architectures and Systems, Volume 1 and 2, Prentice Hall - PTR, NJ, USA, 1999.

[3] D. Abramson, R. Sosic, J. Giddy, and B. Hall, *Nimrod: A Tool for Performing Parametised Simulations using Distributed Workstations*, The 4th IEEE Symposium on High Performance Distributed Computing, Virginia, August 1995.

[4] D. Abramson, J. Giddy, and L. Kotler, *High Performance Parametric Modeling with Nimrod/G: Killer Application for the Global Grid?* International Parallel and Distributed Processing Symposium (IPDPS), IEEE Computer Society Press, 2000.

[5] R. Buyya, D. Abramson and J. Giddy, *Nimrod/G: An Architecture for a Resource Management and Scheduling System in a Global Computational Grid*, 4th Intl. Conf. on High Performance Computing in Asia-Pacific Region (HPC Asia 2000), China.

[6] R. Buyya, D. Abramson and J. Giddy, *Economy Driven Resource Management Architecture for Computational Power Grids*, Intl. Conf. on Parallel and Distributed Processing Techniques and Applications (PDPTA 2000), USA.

[7] R. Buyya, J. Giddy, D. Abramson, *An Evaluation of Economy-based Resource Trading and Scheduling on Computational Power Grids for Parameter Sweep Applications*, The Second Workshop on Active Middleware Services (AMS 2000), In conjunction with HPDC 2001, August 1, 2000, Pittsburgh, USA (Kluwer Academic Press).

[8] R. Buyya, H. Stockinger, J. Giddy, and D. Abramson, *Economic Models for Management of Resources in Peer-to-Peer and Grid Computing*, SPIE International Conference on Commercial Applications for High-Performance Computing, August 20-24, 2001, Denver, USA.

[9] I. Foster and C. Kesselman, Globus: A Metacomputing Infrastructure Toolkit, *International Journal of Supercomputer Applications*, 11(2): 115-128, 1997.

[10] T. Ewing (editor), *DOCK Version 4.0 Reference Manual*, University of California at San Francisco (UCSF), USA, 1998. Online version: http://www.cmpharm.ucsf.edu/kuntz/dock.html

[11] E. Lunney, *Computing in Drug Discovery: The Design Phase*, IEEE Computing in Science and Engineering Magazine, http://computer.org/cise/homepage/2001/05Ind/05ind.htm

[12] Tripos, Inc., SYBYL Mol2 File Format, http://www.tripos.com/services/mol2/, USA, Oct. 7, 2001.

[13] I. Kuntz, J. Blaney, S. Oatley, R. Langridge, and T. Ferrin, *A geometric approach to macromolecule-ligand interactions*, Journal of Molecular Biology, 161: 269-288, 1982.

[14] B. Shoichet, D. Bodian, and I. Kuntz, *Molecular docking using shape descriptors*, Journal of Comp. Chemistry, 13(3): 380-397, 1992.

[15] E. Meng, D. Gschwend, J. Blaney, and I. Kuntz, *Orientational sampling and rigid-body minimization in molecular docking*, Proteins, 17(3): 266-278, 1993.

[16] TurboLinux, *EnFuzion Manual*, http://www.turbolinux.com/downloads/enf/man/enfuzion.htm

[17] R. Buyya, World Wide Grid (WWG), http://www.csse.monash.edu.au/~rajkumar/ecogrid/wwg/